\documentclass[english,prl, twocolumn]{revtex4}

\usepackage[latin1]{inputenc}
\usepackage{graphicx}
\usepackage{amssymb}

\makeatletter

\usepackage{babel}
\makeatother
\begin{document}

\title{Efficient on-chip source of microwave photon pairs in superconducting
circuit QED}

\author{Florian Marquardt}

\affiliation{Physics Department, Center for NanoScience, and Arnold Sommerfeld
Center for Theoretical Physics, Ludwig-Maximilians University Munich,
Theresienstr. 37, 80333 Munich, Germany}

\begin{abstract}
We describe a scheme for the efficient generation of microwave photon
pairs by parametric down-conversion in a superconducting transmission
line resonator coupled to a Cooper pair box serving as an artificial
atom. By properly tuning the first three levels with respect to the
cavity modes, the down-conversion probability may become higher than
in the most efficient schemes for optical photons. We show this by
numerically simulating the dissipative quantum dynamics of the coupled
cavity-box system and discuss the effects of dephasing and relaxation.
The setup analyzed here might form the basis for a future on-chip
source of entangled microwave photons. 
\end{abstract}
\maketitle
\emph{Introduction}. - The generation of photon pairs by parametric
down conversion (PDC)\cite{1970_07_BurnhamWeinberg_PDC_Original,1986_01_HongMandel_SinglePhotonFromPDC,1988_12_ShihAlley_FirstPDC_Bell,1992_BrendelMohlerMartienssen_EnergyTime,1995_12_KwiatEtAl_PDC}
represents one of the basic ways to create nonclassical states of
the electromagnetic field, which has found numerous applications so
far. The conditional detection of one of the photons enables the production
of single photon Fock states \cite{1986_01_HongMandel_SinglePhotonFromPDC,2001_07_WignerFunctionSinglePhoton}.
Furthermore, PDC is the primary method to generate entangled pairs
of particles. Apart from the possibility of testing Bell inequalities
\cite{1988_12_ShihAlley_FirstPDC_Bell,1992_BrendelMohlerMartienssen_EnergyTime,1994_10_TapsterRarityOwen_FransonLongDistance,1995_12_KwiatEtAl_PDC},
this represents a crucial ingredient for a multitude of applications
in the field of quantum information science, ranging from quantum
teleportation through quantum dense coding to quantum key distribution
\cite{2000_BennettDiVincenco_QCReview_Nature}. 

With the advent of superconducting circuit quantum electrodynamics\cite{2004_02_BlaisEtAl_CavityProposal,2004_09_WallraffEtAl_MicrowaveCavity},
it will now be possible to take over many of the concepts that have
been successful in the field of quantum/atom optics and transfer them
to the domain of microwave photons guided along coplanar waveguides
on a chip, interacting with superconducting qubits \cite{1998_01_Devoret_CooperPairBox,1999_04_Nakamura_CooperPairBox,1999_08_Mooij_QubitProposal,2001_04_MakhlinShnirmanSchoen_RMP}.
Recent experiments have realized the strong-coupling limit of the
Jaynes-Cummings model known in atom optics, employing a superconducting
qubit as an artificial two-level atom, and coupling it resonantly
to a harmonic oscillator (i.e. a cavity mode\cite{2004_09_WallraffEtAl_MicrowaveCavity}
or a SQUID \cite{2004_09_MooijFluxQubitJC}). Dispersive QND measurements
of the qubit state, Rabi oscillations and Ramsey fringes have been
demonstrated \cite{2005_04_SchusterWallraff_PRL_DephasingByReadout,2005_08_Wallraff_PRL_UnitVisibility},
leading to a fairly detailed quantitative understanding of the system,
which behaves almost ideally as predicted by theory\cite{2004_02_BlaisEtAl_CavityProposal}. 

\begin{figure}
\includegraphics[%
  width=1\columnwidth]{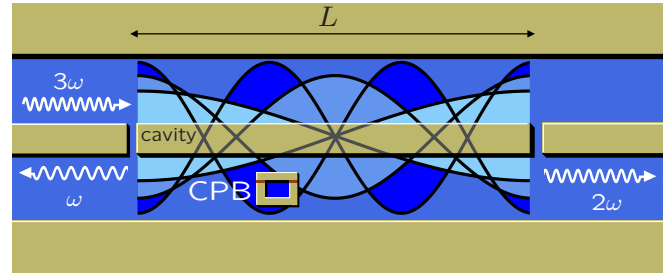}

\caption{\label{Setup}(color online) Schematic setup for the proposed parametric-down
conversion (PDC) experiment in superconducting circuit cavity electrodynamics,
with a Cooper-pair box (CPB) interacting with the three lowest modes
of a transmission line resonator, whose voltage distributions are
shown. }
\end{figure}

\begin{figure}
\includegraphics[%
  width=1\columnwidth]{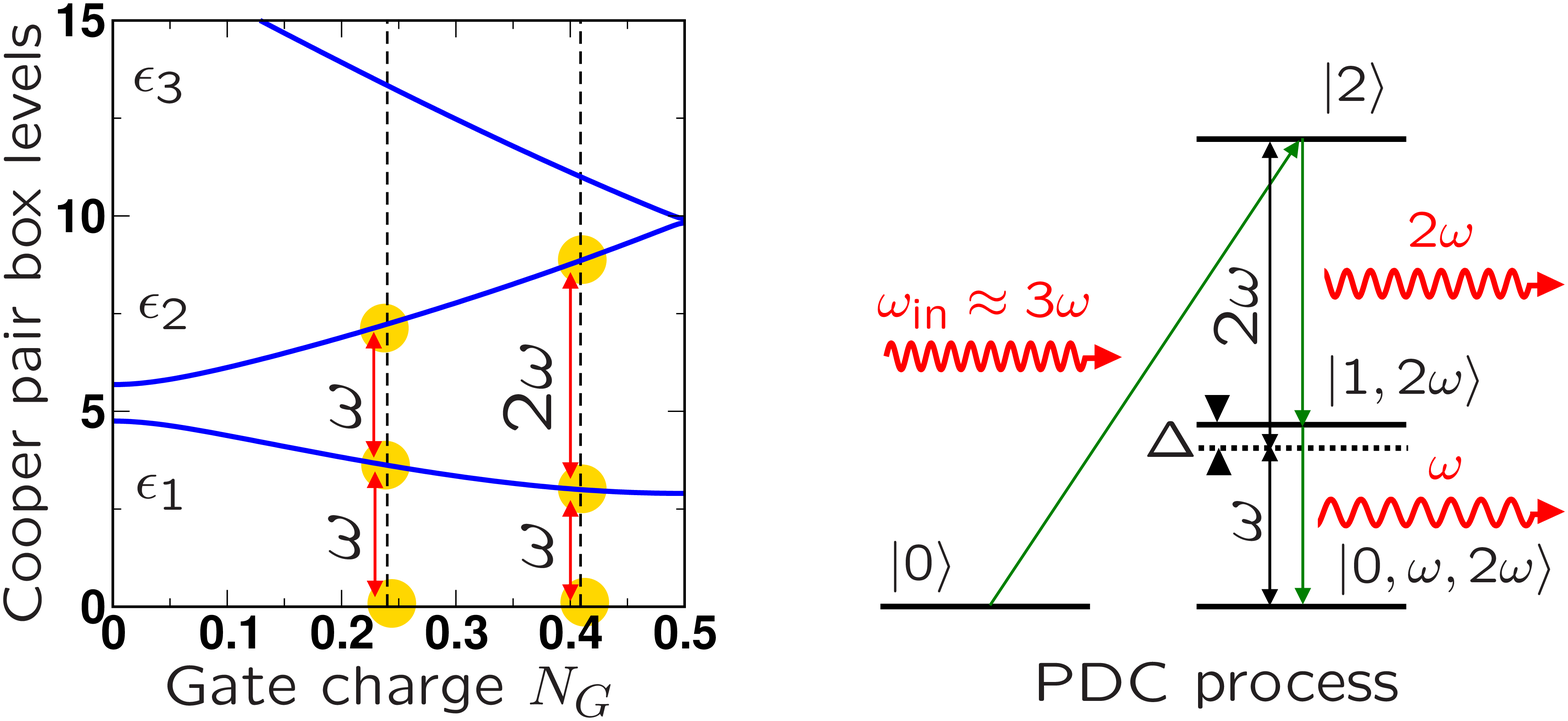}

\caption{\label{cap:Left:-Energy-level-diagram}(color online) Left: Energy-level
diagram of the CPB - transition energies in units of $E_{C}$, for
$E_{J}/E_{C}=3$, as a function of gate charge $N_{G}$. At particular
gate values (away from the {}``qubit'' regime near $N_{G}=1/2$),
the CPB transitions frequencies are related in an integer ratio, $(\epsilon_{2}-\epsilon_{1})/\epsilon_{1}=1:1$
or $2:1$, respectively. This enables to match them with the cavity
modes, giving rise to particularly efficient degenerate or non-degenerate
PDC ($2\omega\mapsto\omega+\omega$ or $3\omega\mapsto\omega+2\omega$,
respectively). Right: Simplified transition scheme for the non-degenerate
PDC process considered in the text, with a detuning $\Delta$ of the
intermediate state $\left|1,2\omega\right\rangle $: $\epsilon_{1}=\omega+\Delta$.}
\end{figure}

In this paper, we will analyze a scheme that implements parametric
down-conversion of microwave photons entering a transmission line
resonator coupled to a Cooper pair box (CPB) providing the required
nonlinearity (Fig. \ref{Setup}). This represents the limit of a single
artificial atom taking the place of the nonlinear crystal usually
employed in optical PDC experiments \cite{1970_07_BurnhamWeinberg_PDC_Original,1986_01_HongMandel_SinglePhotonFromPDC,1988_12_ShihAlley_FirstPDC_Bell,1995_12_KwiatEtAl_PDC},
with the cavity enhancing the PDC rate (cf. \cite{2000_09_OberparleiterWeinfurter_PDC_cavityEnhanced}).
In contrast to other solid state PDC proposals \cite{2000_03_Yamamoto_EntangledPhotonsOnDemand,2002_05_GywatBurkardLoss_EntangledPhotons,2005_09_EmaryMicrowavePhotons},
both the basic cavity setup and the possibility of ejecting the generated
photons into single-mode transmission lines with a high degree of
reliability are already an experimentally proven reality\cite{2004_09_WallraffEtAl_MicrowaveCavity,2005_04_SchusterWallraff_PRL_DephasingByReadout,2005_08_Wallraff_PRL_UnitVisibility}.
Recently, squeezing and degenerate parametric down-conversion have
been analyzed theoretically \cite{2005_06_MoonGirvin_PDCSqueezing}
for a circuit QED setup coupling a charge qubit to two cavity modes.
However, unlike the experiments and most of the theoretical investigations
mentioned above, in this paper we propose to go beyond the regime
where the box may be regarded as a two-level system (qubit), making
use of its first three levels. By further employing its advantage
over real atoms, namely its tunability via the applied magnetic flux
and the gate voltage, this enables us to bring the transitions between
the first three box levels into (near) resonance with the first three
cavity modes (Fig. \ref{cap:Left:-Energy-level-diagram}), thereby
drastically enhancing the resulting probability of (nondegenerate)
parametric down-conversion $\left|3\omega\right\rangle \mapsto\left|\omega\right\rangle \otimes\left|2\omega\right\rangle $.
This represents the major advantage of the present scheme. We treat
the full quantum dissipative dynamics of the box-cavity system, incorporating
the radiation of photons from the cavity as well as nonradiative decay
processes and dephasing in the CPB. We will present results for the
down-conversion efficiency, discuss the minimization of unwanted loss
processes, and comment on possible applications in the end.

\emph{The model}. - The CPB is a device \cite{2001_04_MakhlinShnirmanSchoen_RMP}
in which Cooper pairs can tunnel between two superconducting islands
due to a Josephson-coupling $E_{J}$ (tunable by an external magnetic
flux in a split-junction geometry). The number of transferred Cooper
pairs $\hat{N}$ determines the charging energy, whose scale $E_{C}=e^{2}/2C_{\Sigma}$
is set by the total capacitance of the box island, and which can be
controlled by application of an external gate voltage (expressed in
terms of a gate charge $N_{G}$): \begin{eqnarray}
\hat{H}_{{\rm CPB}} & = & 4E_{C}(\hat{N}-N_{G})^{2}\nonumber \\
 &  & -\frac{E_{J}}{2}\sum_{N}\left|N+1\right\rangle _{c}\left\langle N\right|_{c}+{\rm h.c.}\label{HamCPB}\end{eqnarray}
Here $\left|N\right\rangle _{c}$ represents a charge state of the
CPB. Two gaps in a superconducting coplanar waveguide act as mirrors
of a cavity for microwave photons, $\hat{H}_{{\rm cavity}}=\sum_{j=1}^{3}\omega_{j}\hat{a}_{j}^{\dagger}\hat{a}_{j}$,
where we will focus attention on the three lowest-lying cavity modes
with $\omega_{j}=j\omega$ (we set $\hbar\equiv1$). The electric
field inside the cavity acts on the CPB via a capacitive coupling,
adding a fluctuating quantum-mechanical component to the gate charge
$N_{G}$. This results \cite{2004_02_BlaisEtAl_CavityProposal} in
an interaction 

\begin{equation}
\hat{H}_{{\rm int}}=\left[\sum_{j=1}^{3}g_{j}(\hat{a}_{j}+\hat{a}_{j}^{\dagger})\right]\hat{N}\,.\label{Hint}\end{equation}
The coupling constants (vacuum Rabi frequencies) $g_{j}=g_{0}\sqrt{j}\phi_{j}(x)$,
with $g_{0}=2\frac{eC_{g}}{C_{\Sigma}}\sqrt{\omega/Lc}$, are given
in terms of the mode functions $\phi_{1}(x)=\sin(x\pi/L)$, $\phi_{2}(x)=\cos(2\pi x/L)$
and $\phi_{3}(x)=\sin(3\pi x/L)$, which are defined on the interval
$x=-L/2\ldots L/2$. 

The full Hamiltonian forming the basis of our analysis is thus given
by

\begin{equation}
\hat{H}=\hat{H}_{{\rm cavity}}+\hat{H}_{{\rm CPB}}+\hat{H}_{{\rm int}}+\hat{H}_{{\rm env}}=\hat{H}_{0}+\hat{H}_{{\rm env}}\,,\label{HamiltonianSum}\end{equation}
where $\hat{H}_{{\rm env}}$ includes the coupling to the various
reservoirs forming the environment. This involves both the possibility
for microwave photons to leak out of the cavity, as well as the various
possible nonradiative decay and decoherence processes acting on the
CPB. The details will be specified below when setting up the master
equation.

\emph{Basic considerations}. - At least three basic features distinguish
such a setup from the usual optical photon PDC experiments employing
nonlinear crystals: (i) There is no momentum conservation, as the
system is essentially zero-dimensional. (ii) However, energy conservation
is much more restrictive, as the set of possible frequencies is limited
to the discrete cavity modes. For appropriate input frequency, this
results in a resonant enhancement of the PDC process. In contrast
to a passive filtering scheme, the bandwidth of the generated photons
is reduced without diminishing the signal intensity. (iii) The microwave
polarization is fixed and thus cannot be used for entanglement. At
the end of this paper we will point out other options that can be
explored. 

\emph{Estimating the PDC rate}. - If the CPB is operated as a two-level
system (qubit) \cite{2005_06_MoonGirvin_PDCSqueezing}, the decay
of a $3\omega$ photon into two lower-energy photons comes about through
multi-step transitions, where at least one of the intermediate virtual
states will have an energy detuning of the order of $\omega$, which
contributes a small factor $(g/\omega)^{2}$ to the PDC rate.

However, we can enhance the PDC rate by exploiting more than the first
two levels of the CPB. Indeed, by properly tuning the Josephson coupling
$E_{J}$ and the gate charge $N_{G}$, it is possible to make the
transitions between the first three CPB energy levels $\left|0\right\rangle ,\,\left|1\right\rangle ,\,\left|2\right\rangle $
resonant with the cavity modes. The PDC process we want to consider
is thus $\left|0,3\omega\right\rangle \mapsto\left|2\right\rangle \mapsto\left|1,2\omega\right\rangle \mapsto\left|0,\omega,2\omega\right\rangle $,
with $\epsilon_{2}\approx3\omega$ and $\epsilon_{1}=\omega+\Delta$,
such that all the transitions are (nearly) resonant. This reduces
the largest energy denominator to the detuning $\Delta$, resulting
in an enhancement of the PDC rate by a factor of $(\omega/\Delta)^{2}$.
What limits the enhancement? If $\Delta$ is made too small, one may
end up with less than a complete pair of downconverted photons: Instead
of passing virtually through the intermediate state $\left|1,2\omega\right\rangle $
containing a $2\omega$-photon and the CPB in its first excited level,
that state will acquire a significant real population. As a result,
the temporal correlation between photons would be destroyed, and nonradiative
decays $\left|1,2\omega\right\rangle \mapsto\left|0,2\omega\right\rangle $
may occur, without emitting the second photon of frequency $\omega$.
Clearly, there is a tradeoff between the achieved PDC probability
and the fidelity of down-conversion. This will be confirmed by the
detailed simulations below.

We will find that it is possible to achieve a PDC probability in the
percent range that surpasses that of the most efficient modern optical
PDC schemes\cite{2004_09_Edamatsu_UV_PDC_Semiconductor} (which have
a PDC probability of about $10^{-4}$; though the absolute PDC \emph{rate}
in those experiments is about $10^{9}$ times larger due to the drastically
higher input power). Earlier well-known optical PDC experiments\cite{1995_12_KwiatEtAl_PDC}
generated less than one usable coincidence detection event for every
$10^{13}$ incoming photons.

\emph{Simulation of the quantum-dissipative dynamics}. - In the ideal
case, one could integrate out the intermediate state, yielding an
effective PDC term of the form $\left|0\right\rangle \left\langle 2\right|\hat{a}_{1}^{\dagger}\hat{a}_{2}^{\dagger}+h.c.$.
However, here we take into account all loss processes, by solving
for the full dynamics of the CPB/cavity-system under an external microwave
drive of frequency $\omega_{{\rm in}}\approx3\omega$, using a Markoff
master equation of Lindblad form: \begin{equation}
\frac{d\hat{\rho}}{dt}=(\mathcal{L}_{{\rm 0}}+\mathcal{L}_{{\rm drive}}+\mathcal{L}_{{\rm cavity}}^{{\rm decay}}+\mathcal{L}_{{\rm CPB}}^{{\rm relax}}+\mathcal{L}_{{\rm CPB}}^{{\rm deph}})\hat{\rho}\label{MEQ}\end{equation}
Here $\mathcal{L}_{{\rm 0}}\hat{\rho}=-i[\hat{H}_{0},\hat{\rho}]$,
and the external microwave input, at a frequency $\omega_{{\rm in}}\approx3\omega$
and with an amplitude $\alpha$, is described by $\mathcal{L}_{{\rm drive}}\hat{\rho}=-i[\hat{H}_{{\rm drive}}(t),\hat{\rho}]$,
with $\hat{H}_{{\rm drive}}(t)=\alpha\hat{a}_{3}^{\dagger}e^{-i\omega_{{\rm in}}t}+{\rm h.c.}$.

The dissipative terms in the Liouvillian are of Lindblad form

\begin{equation}
\mathcal{L}[\hat{A}]\hat{\rho}\equiv\hat{A}\hat{\rho}\hat{A}^{\dagger}-\frac{1}{2}\hat{A}^{\dagger}\hat{A}\hat{\rho}-\frac{1}{2}\hat{\rho}\hat{A}^{\dagger}\hat{A}\,.\end{equation}
 They describe: the decay of each cavity mode at a rate $\kappa_{j}$
($j=1,2,3$),

\begin{equation}
\mathcal{L}_{{\rm cavity}}^{{\rm decay}}=\sum_{j}\kappa_{j}\,\mathcal{L}[\hat{a}_{j}],\label{Lcavdecay}\end{equation}
pure dephasing processes in the CPB that do not lead to transitions
between levels (at rates $\gamma_{\varphi,j}$),

\begin{equation}
\mathcal{L}_{{\rm CPB}}^{{\rm deph}}=\sum_{j}\gamma_{\varphi,j}\mathcal{L}[\left|j\right\rangle \left\langle j\right|],\label{LCPBdeph}\end{equation}
 and nonradiative relaxation processes leading from a level $l$ to
a lower energy level $j$ of the qubit: 

\begin{equation}
\mathcal{L}_{{\rm CPB}}^{{\rm relax}}=\sum_{j<l}\gamma_{j\leftarrow l}\mathcal{L}[\left|j\right\rangle \left\langle l\right|]\label{LCPBrelax}\end{equation}

We keep only resonant terms ({}``rotating wave approximation'',
RWA) in the CPB-cavity interaction and go over to a frame rotating
at $\omega_{{\rm in}}$, which eliminates the time-dependence in $\hat{H}_{{\rm drive}}$,
but replaces $\hat{H}_{0}$ by $\hat{H}_{0}-\hat{W}$, with $3\hat{W}/\omega_{{\rm in}}=\sum_{j}j\hat{a}_{j}^{\dagger}\hat{a}_{j}+\left|1\right\rangle \left\langle 1\right|+3\left|2\right\rangle \left\langle 2\right|$.
This is accomplished by applying the time-dependent unitary transformation
$\exp(i\hat{W}t)\hat{A}\exp(-i\hat{W}t)$ to the Hamiltonian $\hat{H}_{0}+\hat{H}_{{\rm drive}}$
and the density matrix. 

\begin{figure}
\includegraphics[%
  width=0.8\columnwidth]{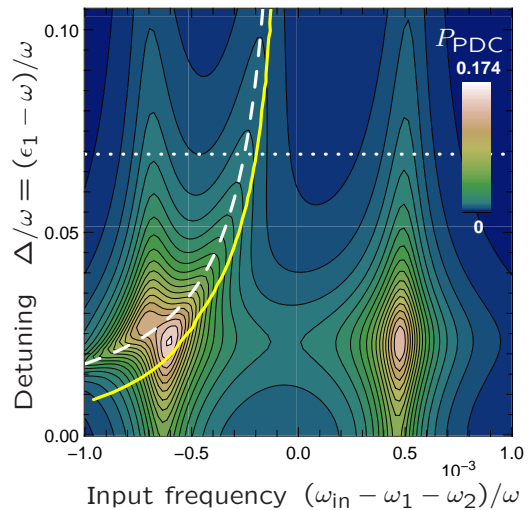}

\caption{\label{freqsweep}(color online) Parametric down-conversion probability
$P_{{\rm PDC}}=P(\left|\omega,2\omega\right\rangle )\kappa^{2}/|\alpha|^{2}$,
as a function of the microwave input frequency $\omega_{{\rm in}}$
and the detuning of the intermediate state. The dashed line indicates
the analytical resonance condition (see main text), the full line
denotes the location of minimal $Q_{1}$ (high fidelity of PDC), and
the dotted line is the cross-section shown in Fig. \ref{Freq2}.}
\end{figure}

We have obtained numerical solutions of the master equation for a
wide range of parameters, and the results are shown in Figs. \ref{freqsweep}
and \ref{twoDplot}. All these simulations have been performed in
a Hilbert space that has been truncated under the assumption of a
small external drive (The maximum excitation energy of the qubit+cavity
system is restricted to $3\omega$, and the down-conversion rate is
linear in the input power). For the plots shown below, we have used
experimentally reasonable parameter values: $g_{0}/\omega=10^{-2},\,\kappa_{j}/\omega=10^{-4},\,\gamma_{j\leftarrow l}/\omega=10^{-5}\,(j<l),\,\gamma_{\varphi,j}/\omega=2\cdot10^{-4}\,(j>0)$.
For reference, note that $\omega\sim10\,{\rm GHz}$ in typical experiments\cite{2004_09_WallraffEtAl_MicrowaveCavity}.
We have placed the CPB at $x=0.3\times L$.

\emph{Discussion}. - In order to interpret the results, we note that
the production of photon pairs at a rate $\Gamma_{{\rm PDC}}$ is
balanced by the decay of photons out of the cavity, at a rate $\kappa$.
Thus, in an ideal lossless cavity PDC scheme, the probabilities to
find the cavity in the states $\left|\omega,2\omega\right\rangle $,
$\left|\omega\right\rangle $ and $\left|2\omega\right\rangle $ all
become equal to $\Gamma_{{\rm PDC}}/(2\kappa)$. Therefore, we define
$\Gamma_{{\rm PDC}}/(2\kappa)\equiv P_{\left|\omega,2\omega\right\rangle }$.
The rate $R$ of incoming photons is given by $R=2|\alpha|^{2}/\kappa$.
Thus, the PDC probability (chance of a given photon undergoing PDC)
can be set to $P_{{\rm PDC}}=\Gamma_{{\rm PDC}}/R=P_{\left|\omega,2\omega\right\rangle }\kappa^{2}/|\alpha|^{2}$
(we will use this as a definition even where the scheme deviates from
ideal conditions, discussing the PDC fidelity separately). %

\begin{figure}
\includegraphics[%
  width=1\columnwidth]{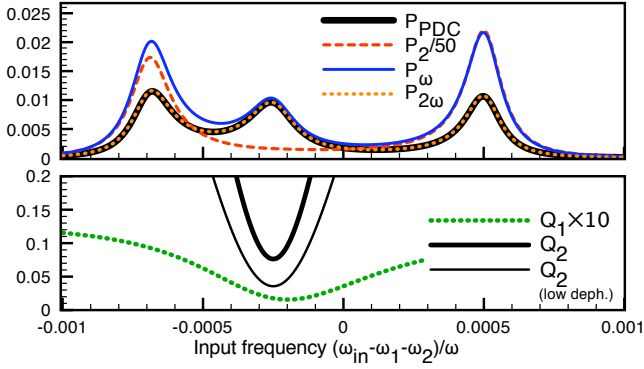}

\caption{\label{Freq2}(color online) \emph{Top}: The parametric down conversion
probability as a function of input frequency along the cross-section
indicated in Fig. \ref{freqsweep}, and the probabilities $P_{\omega}=fP(\left|\omega\right\rangle )$
and $P_{2\omega}=fP(\left|2\omega\right\rangle )$ for one or two
photons in the cavity, rescaled in the same manner, with $f=\kappa^{2}/|\alpha|^{2}$
(ideally $P_{{\rm PDC}}=P_{\omega}=P_{2\omega}$). The probability
of an excited qubit, $P_{2}=fP(\left|2\omega\right\rangle )$, is
also shown. \emph{Bottom}: The {}``non-ideality parameters'' $Q_{1}=\gamma P_{\left|1,2\omega\right\rangle }/(\kappa P_{\left|\omega,2\omega\right\rangle })$
and $Q_{2}=P_{\left|\omega\right\rangle }/P_{\left|\omega,2\omega\right\rangle }-1$,
reaching a minimum at the middle peak (thin line: for half the dephasing
rate of $\gamma_{\varphi}/\omega=2\cdot10^{-4}$).}
\end{figure}

\begin{figure}
\includegraphics[%
  clip,
  width=1\columnwidth]{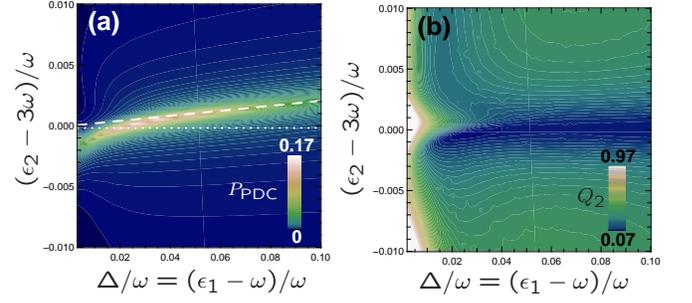}

\caption{\label{twoDplot}(color online) (a) Parametric down-conversion probability
$P{}_{{\rm PDC}}$, and (b) the non-ideality parameter $Q_{2}$, as
a function of the detunings between the Cooper-pair box energy levels
$\epsilon_{1,2}$ and the cavity modes (controlled by $E_{J}$ and
$N_{G}$). At each point, the microwave input frequency $\omega_{{\rm in}}$
has been chosen to minimize the parameter $Q_{1}$ (cf. Fig. \ref{Freq2}).
The dotted line indicates the cross-section displayed in Fig. \ref{freqsweep}.}
\end{figure}

In Fig. \ref{freqsweep}, we have plotted the PDC probability as a
function of the input frequency $\omega_{{\rm in}}$ and the detuning
$\Delta=\epsilon_{1}-\omega$. In general, $P_{{\rm PDC}}$ becomes
maximal when $\omega_{{\rm in}}$ matches the doublet frequencies
$\epsilon_{2}$ and $3\omega$ (modified by the vacuum Rabi splitting):
See the two vertical {}``ridges'', independent of $\Delta$. There
is a third resonance at $\omega_{{\rm in}}=3\omega-\tilde{g}_{1}^{2}/\Delta$,
with a dispersive shift depending on $\Delta$, for which the input
frequency matches the energy of the outgoing state (dashed curve).
Here we have defined $\tilde{g}_{1}=g_{1}\left\langle 1\left|\hat{N}\right|0\right\rangle $
as the actual vacuum Rabi frequency. Although $P_{{\rm PDC}}$ becomes
maximal when any of these resonances cross, this does not yield ideal
photon pairs, as will become clear shortly. Turning to Fig. \ref{Freq2}
(top), we see a cross-section of $P_{{\rm PDC}}$ (as a function of
$\omega_{{\rm in}}$). In addition, we have included the probabilities
of having one or two photons inside the cavity, which (in this normalization)
should be identical to $P_{{\rm PDC}}$ in an ideal scheme. The apparent
surplus in $\omega$-photons is due to a nonradiative process, as
we will discuss now. 

The relevant decay routes are: (i) $\left|2\right\rangle \mapsto\left|1\right\rangle \leftrightarrow\left|\omega\right\rangle \mapsto\left|0\right\rangle $,
(ii) $\left|\omega,2\omega\right\rangle \mapsto\left|\omega\right\rangle \leftrightarrow\left|1\right\rangle \mapsto\left|0\right\rangle $,
(iii) $\left|1,2\omega\right\rangle \mapsto\left|1\right\rangle \mapsto\left|0\right\rangle $,
and (iv) $\left|1,2\omega\right\rangle \mapsto\left|2\omega\right\rangle \mapsto\left|0\right\rangle $.
Process (i) leads to a single $\omega$ photon being emitted, while
(ii)-(iv) produce a single $2\omega$ photon with no corresponding
partner photon. All of these processes get suppressed with an increasing
energy mismatch $|\Delta|=|\epsilon_{1}-\omega|$ between the states
$\left|1\right\rangle $ and $\left|\omega\right\rangle $. A larger
broadening of the levels (produced by dephasing or decay) may partially
overcome this energy mismatch, leading to a higher rate of unwanted
loss processes. In order to quantify these processes, we have plotted,
in Fig. \ref{Freq2} (bottom), the {}``non-ideality measures'' $Q_{1}=\gamma P_{\left|1,2\omega\right\rangle }/(\kappa P_{\left|\omega,2\omega\right\rangle })$
and $Q_{2}=P_{\left|\omega\right\rangle }/P_{\left|\omega,2\omega\right\rangle }-1$
. Here $Q_{1}$ measures the ratio of nonradiative relaxation from
the intermediate state $\left|1,2\omega\right\rangle $ to the rate
of pair emission, while $Q_{2}$ indicates the degree to which the
populations of $\left|\omega\right\rangle $ and $\left|\omega,2\omega\right\rangle $
are identical (which is the case ideally, for $\kappa_{j}\equiv\kappa$).
Both $Q_{1}$ and $\left|Q_{2}\right|$ should be as small as possible.
The doublet peaks yield a large PDC rate, but also a large population
of the CPB excited state $\left|2\right\rangle $, leading to the
decay process (i) and a resulting surplus of $\omega$-photons (Fig.
\ref{Freq2}, top). Thus, we observe that the vacuum Rabi splitting
of the doublet is essential: it allows for the appearance of the third
(middle) peak in $P_{{\rm PDC}}$ that has a far lower qubit population
and corresponding rate of unwanted loss processes (minima in $Q_{1,2}$).
Any reduction in the broadening of the peaks (set by $\kappa,\gamma,\gamma_{\varphi}$)
helps to further increase the quality of PDC. 

The PDC quality can also be measured by evaluating the 2-photon correlator
$K_{jl}(t)=\left\langle \hat{a}_{l}^{\dagger}\hat{a}_{j}^{\dagger}(t)\hat{a}_{j}(t)\hat{a}_{l}\right\rangle /\left(\left\langle \hat{a}_{l}^{\dagger}\hat{a}_{l}\right\rangle \left\langle \hat{a}_{j}^{\dagger}\hat{a}_{j}\right\rangle \right),$
which determines the probability to detect a mode $j$ photon at time
$t$ inside the cavity, provided a mode $l$ photon has been detected
at $t=0$. Using the quantum regression theorem applied to our master
equations, we have checked that at small values of $Q_{1,2}$, the
ideal case is indeed observed, where the correlator decays at a rate
$\kappa$, both for $(j,l)=(1,2)$ and $(j,l)=(2,1)$, while nonradiative
processes change these decay rates and make them unequal. 

The gate charge $N_{G}$ and the Josephson coupling $E_{J}$ change
the relevant energies $\epsilon_{1}$ and $\epsilon_{2}$ of the first
two excited states of the Cooper pair box, and, to a lesser degree,
the matrix elements for the coupling to the cavity field. At a given
value of $E_{J}$, one can tune to $N_{G}=N_{G}^{*}[E_{J}]$, such
that the bare resonance condition $\epsilon_{2}=3\omega$ is fulfilled.
In the following, we consider the effects of a small additional {}``offset
gate charge'' $\delta N_{G}$, which mainly changes $\epsilon_{2}$,
while $E_{J}$ itself is used to tune $\epsilon_{1}$. The plots discussed
so far have been obtained at fixed $\delta N_{G}$, while changing
$E_{J}$. At any given value of $E_{J}$ and $\delta N_{G}$, it is
possible to select an input frequency $\omega_{{\rm in}}$ which minimizes
$Q_{1}$. The resulting PDC rate has been plotted as a function of
the detunings of the two CPB energy levels, $\epsilon_{1}$ and $\epsilon_{2}$,
in Fig. \ref{twoDplot} (a), while the corresponding parameter $Q_{2}$
is shown in Fig. \ref{twoDplot} (b) ($Q_{1}$ is below $0.01$ in
the relevant region where $Q_{2}$ is small). 

We conclude that, while maintaining a good reliability of the PDC
process, the down conversion probability can become on the order of
a few percent in the present setup, which thus indeed represents a
highly efficient source of photon pairs. Similar results have been
found for other reasonable parameter sets (e.g. $g_{0}/\omega=10^{-2},\,\kappa_{j}/\omega=10^{-5},\,\gamma_{j\leftarrow l}/\omega=10^{-4}\,(j<l),\,\gamma_{\varphi,j}/\omega=10^{-3}\,(j>0)$).

\emph{Generation of entanglement.} - The down-converted $\omega-$
and $2\omega$-photons can independently leak out of either side of
the cavity. By post-selecting (cf. \cite{1982_09_Aspect_SwitchingPolarizers,1988_12_ShihAlley_FirstPDC_Bell})
only events where a photon is detected both in the left and the right
arm each, one ends up with a frequency-entangled state that is directly
equivalent to the entangled triplet state: $\left|2\omega\right\rangle _{L}\otimes\left|\omega\right\rangle _{R}+\left|\omega\right\rangle _{L}\otimes\left|2\omega\right\rangle _{R}\,.$We
note, however, that a full Bell test requires measurements in a superposition
basis, which is hard to realize for states of different energies.

Another, simpler, possibility is to test for energy-time entanglement
as first proposed by Franson \cite{1989_05_Franson_BI_EnergyTime,1992_BrendelMohlerMartienssen_EnergyTime,1994_10_TapsterRarityOwen_FransonLongDistance}.
This requires feeding the generated photons into Mach-Zehnder interferometers,
each of them containing a short and a long arm as well as a variable
phase-shifter in one of the arms. By measuring the photon-detection
correlation between the altogether four output ports of the two interferometers,
it is possible to violate the usual kinds of Bell inequalities. The
great advantage of such a scheme (particularly in the context of superconducting
circuit QED) is that it does not require the polarization as a degree
of freedom. 

A less demanding, first experimental test of the PDC source described
here might measure the intensity cross-correlations of the microwave
output beams (at $\omega$ and $2\omega$) or implement homodyning
techniques \cite{2001_07_WignerFunctionSinglePhoton,2005_09_Gross_MicrowavePhotonsFluxQubit}
to characterize the quantum state. Finally, it is worth noting that
for $\omega_{3}\approx\epsilon_{2}$ a sufficiently strong vacuum
Rabi splitting between $\left|3\omega\right\rangle $ and $\left|2\right\rangle $
in principle enables a scheme where a Rabi $\pi$ pulse is used to
put exactly one excitation into the system, which then decays in the
way described here, thus realizing a source of microwave photon pairs
on demand.

\emph{Conclusions}. - In this paper we have described and analyzed
a setup for parametric down conversion in superconducting circuit
cavity QED, suitable for the generation of pairs of entangled microwave
photons. In contrast to earlier discussions, we have considered employing
a transition via the first three levels of the artificial atom (Cooper
pair box), which can be tuned to achieve a drastically enhanced PDC
rate. We have analyzed the trade-off between optimizing the PDC rate
and minimizing loss processes, by carrying out extensive numerical
simulations of the quantum dissipative dynamics. The setup described
here can be realized by moderate modifications of existing experiments,
and it can hopefully form the basis for more detailed investigations
into the nonclassical properties of the microwave field in circuit
QED experiments.

\emph{Acknowledgments}. I thank S. Girvin, A. Wallraff, A. Blais,
J. Majer, D. Schuster, M. Mariantoni, E. Solano and R. Schoelkopf
for discussions, and especially M. H. Devoret for pointing out the
potential use of a three-level configuration. This work was supported
by the DFG.

\bibliographystyle{apsrev}
\bibliography{/home/florian/pre/bib/shortall}

\end{document}